\documentclass{ifacconf}

\usepackage{graphicx}      
\usepackage{epstopdf} 
\usepackage[round]{natbib} 
\usepackage{tabularx}
\usepackage{dblfloatfix}
\usepackage{comment}
\usepackage{makecell}
\usepackage{placeins}
\usepackage{
    physics, 
    siunitx, 
    amsmath, 
    amssymb, 
    mathtools, 
    amsfonts, 
    booktabs,
    tabularx,
    subcaption
}

\newcommand{\ts}[1]{_{\text{#1}}}

\newcommand{\mybar}{\overline}

\begin{document}
\begin{frontmatter}

\title{Bayesian Parameter Estimation Applied to the Li-ion Battery Single Particle Model with Electrolyte Dynamics  \thanksref{footnoteinfo} \thanksref{finfo2}} 

\thanks[footnoteinfo]{This work was carried out with funding  received from the Faraday Institution ({\tt faraday.ac.uk}; EP/S003053/1,  ref.\ FIRG003). Scott Marquis was supported by the EPSRC Centre For Doctoral Training in Industrially Focused Mathematical Modelling (EP/L015803/1) in collaboration with Siemens Corporate Technology.}

\thanks[finfo2]{This work has been submitted to IFAC for possible publication.}

\author[First]{Antti Aitio}, 
\author[Second]{Scott G. Marquis},
\author[First]{Pedro Ascencio},
\author[First]{David Howey}

\address[First]{Department of Engineering Science, University of Oxford, Oxford OX1 3PJ, United Kingdom (e-mails: {\tt antti.aitio@eng.ox.ac.uk, pedro.ascencio@eng.ox.ac.uk, david.howey@eng.ox.ac.uk}).}
\address[Second]{Mathematical Institute, University of Oxford, OX2 6GG, United Kingdom (e-mail:  {\tt marquis@maths.ox.ac.uk})}

\begin{abstract}                
This paper presents a Bayesian parameter estimation approach and identifiability analysis for a lithium-ion battery model, to determine the uniqueness, evaluate the sensitivity and quantify the uncertainty of a subset of the model parameters. The analysis was based on the single particle model with electrolyte dynamics, rigorously derived from the Doyle-Fuller-Newman model using  asymptotic analysis including electrode-average terms. The Bayesian approach allows complex target distributions to be estimated, which enables a global analysis of the parameter space. The analysis focuses on the identification problem (i) locally, under a set of discrete quasi-steady  states of charge, and in comparison (ii) globally with a continuous excursion of state of charge. The performance of the methodology was evaluated using synthetic data from multiple numerical simulations under diverse types of current excitation. We show that various diffusivities as well as the transference number may be estimated with small variances in the global case, but with much larger uncertainty in the local estimation case. This also has significant implications for estimation where parameters might vary as a function of state of charge or other latent variables.
\end{abstract}


\end{frontmatter}

\section{Introduction}
Design, estimation and control of lithium-ion batteries requires accurate prediction of performance, and physics-based electrochemical models have been shown to provide this. The so-called single particle model (SPM) is one the simplest models of this type \citep{Moura:17} and, despite its limitations \citep{Chaturvedi:10}, it preserves the fundamental estimation and control challenges \citep{Moura:2015} of the more complex Doyle-Fuller-Newman (DFN) model \citep{Doyle:93}. One of these challenges, addressed here, is the identification of the parameters involved in the coupled infinite-dimensional formulation of the battery internal dynamics, including mass transport, interfacial reaction kinetics and ohmic losses.

Several authors have attempted to address the battery parameter identification problem \citep{Forman:2012,Zhang:2018,Chun:2019} with varying results, and only a few studies in this space have begun to consider a Bayesian approach \citep{Sethurajan:2018,Lopez:2016,Ramadesigan:2011}. A relevant aspect of Li-ion batteries is the functional dependence of the model parameters with respect to the internal states \citep{Ding:2001,Jossen:2006,Ng:2020}. Commonly, the parameter estimation problem is performed by maximum likelihood estimation (MLE), interpolating \emph{local} estimates at a number of points, for instance at different states of charge (SOC), via electrochemical impedance spectroscopy (EIS) or galvanostatic intermittent titration technique (GITT)\citep{Westerhoff:2016,Ecker_01:2015}. To quantify the uncertainty of parameters estimated with this approach, typically the Fisher information criterion is used \citep{Santhanagopalan:2007, Schmidt:2010, Lin:2015}. 

In this paper, in contrast to the standard MLE approach, we introduce a comprehensive Bayesian parameter estimation technique. This approach enables us to determine the uniqueness of the mode of the posterior parameter distribution, and to evaluate  the global parameter sensitivity. In order to perform this study, the single particle model with electrolyte (SPMe) dynamics was used, and it was derived from asymptotic analysis of the Doyle-Fuller-Newman model, including electrode-average terms. To solve the parameter estimation problem, a Monte Carlo technique was implemented and applied to synthetically generated input-output (voltage and current) data, both at a set of discrete quasi-steady states of charge, and over a continuous excursion range.

\section{Model Formulation}
\begin{figure}[h]
\vspace{-0.3cm}%
\centering%
\includegraphics[width=0.48\textwidth]{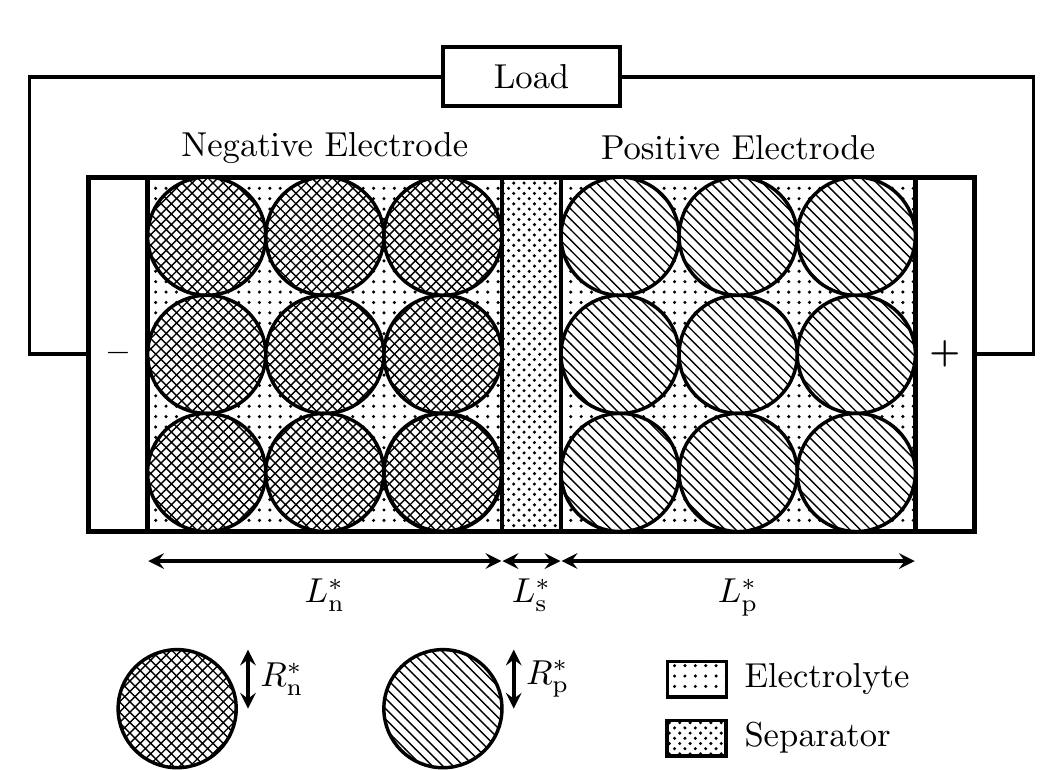}%
\vspace{-0.2cm}%
\caption{Schematic of a lithium-ion battery}%
\label{fig:lion-battery}%
\end{figure}

We consider a single lithium-ion cell with a negative electrode thickness of $L\ts{n}$ with particles of radius $R\ts{n}$, a separator thickness of $L\ts{s}$, and a positive electrode thickness of $L\ts{p}$ with particles of radius $R\ts{p}$, as displayed in Figure \ref{fig:lion-battery}. We also define the total thickness of the cell to be $L=L\ts{n}+L\ts{s}+L\ts{p}$. 

In \cite{Marquis:2019}, we employed asymptotic methods to systematically derive a single particle model with electrolyte (SPMe) from the full DFN model. This enabled the identification of the assumptions in ad-hoc derivations of the SPMe that are appropriate vs.\ those that introduce unnecessary errors. The resulting model was compared to both the standard SPM, and a number of ad-hoc versions of the SPMe from the literature (\cite{perezSPMe, kemperSPMe, mouraSPMe}). It was found that our approach improved the accuracy with negligible additional computational cost.  

We now introduce the asymptotic SPMe and highlight the key distinguishing features of the model. The concentration of lithium-ions in the electrolyte is denoted by $c\ts{e}$ and the concentrations of lithium in the negative and positive electrode particles are denoted $c\ts{s,n}$ and $c\ts{s,p}$, respectively. The variable $x\in[0, L]$ is used to denote through-cell position and $r\ts{k}\in R\ts{k}$ to denote the in-particle radial position. All parameters are defined in Table \ref{tab:parameters}. The single particle in the SPMe represents a theoretical electrode-averaged particle. We highlight this by using an overbar to represent an electrode-averaged quantity. With this notation, the evolution equation for the lithium concentration in the electrode-averaged particle is: 

\vspace{-0.2cm}

\begin{subequations}\label{eqn:SPMe-particles}
\begin{equation}
    \pdv{\mybar{c}\ts{s,k}}{t} = \frac{D\ts{s,k}}{r^2}\pdv{r}\left(r^2 \pdv{\mybar{c}\ts{s,k}}{r}\right),
\end{equation}%
\begin{equation}
    \pdv{\mybar{c}\ts{s,k}}{r}\bigg|_{r=0} = 0, 
\end{equation}%
\begin{equation}
    -D\ts{s,k}\pdv{\mybar{c}\ts{s,k}}{r}\bigg|_{r=R\ts{k}} =
    \begin{cases}
        \frac{I}{F a\ts{n} L\ts{n}}, \quad &\text{k}=\text{n}, \\ 
        -\frac{I}{F a\ts{p} L\ts{p}}, \quad &\text{k}=\text{p}.
    \end{cases} 
\end{equation}
\end{subequations}

\vspace{-0.2cm}

In the electrolyte, the lithium-ion concentration evolves according the the porous-electrode form of the Onsager--Stefan--Maxwell equations. In the SPMe, the source/sink term representing lithium-ion exchange with the electrode active material is just that of the electrode-average lithium-ion exchange (as opposed to the full Butler--Volmer version in the DFN model). Additionally, the SPMe employs the cell-averaged diffusivity thus linearising the equations. Therefore, the lithium-ion concentration in the electrolyte is governed by:

\vspace{-0.2cm}

\begin{subequations}\label{eqn:SPMe-electrolyte}
\begin{equation}
\label{eqn:SPMe-dimensional-electrolyte}
     \epsilon\ts{k}\pdv{c\ts{e}}{t} = \epsilon\ts{k}^b D\ts{e,typ} \pdv[2]{c\ts{e}}{x} +  
    \begin{cases} 
        \frac{(1-t^+)I}{FL\ts{n}}, \quad &\text{k}=\text{n}, \\ 
        0, \quad &\text{k}=\text{s}, \\ 
        -\frac{(1-t^+)I}{FL\ts{p}}, \quad &\text{k}=\text{p},
    \end{cases}
\end{equation}
\begin{equation}
    \pdv{c\ts{e}}{x}\bigg|_{x=0} = \pdv{c\ts{e}}{x}\bigg|_{x=L} = 0
\end{equation}
\begin{equation}
    c\ts{e}(x, 0) = c\ts{e,typ}. 
\end{equation}
\end{subequations}

The terminal voltage, $V$, of a cell is typically written as 
\begin{equation}\label{eqn:pointwise-voltage}
    V = U\ts{eq} + \eta\ts{r} + \eta\ts{c} + \Delta\Phi\ts{Elec} + \Delta \Phi\ts{Solid}
\end{equation}
where $U\ts{eq}$ is the open-circuit voltage, $\eta\ts{r}$ is the reaction overpotentials, $\eta\ts{c}$ is the electrolyte concentration overpotential, $\Delta\Phi\ts{Elec}$ is the electrolyte Ohmic losses, and $\Delta\Phi\ts{Solid}$ is the solid-phase Ohmic losses. In this expression, each term represents the value of that term when the electrochemical reactions occurs at one point/particle in each electrode. At this stage, ad-hoc SPMe models choose a particular point/particle at which to evaluate (\ref{eqn:pointwise-voltage}). However, this cannot be done without introducing an error larger than the other simplification errors. This is because it is not possible to determine the flux into a particular particle, because one cannot also determine the local reaction overpotential which drives the flux. Therefore, the ad-hoc models assume that the flux into the particle is equal to the average flux into all particles in that electrode, which introduces the error. Ad-hoc models also typically choose to evaluate (\ref{eqn:pointwise-voltage}) at the particles nearest to the current collectors, which can experience fluxes far from the average in each electrode. 

To avoid introducing the larger errors associated with evaluating (\ref{eqn:pointwise-voltage}) at a particular point, we average across all particles in each electrode to obtain the terminal voltage in terms of electrode-averaged quantities
\begin{subequations}
\begin{equation}\label{eqn:electrode-averaged-voltage}
    V = \mybar{U}\ts{eq} + \mybar{\eta}\ts{r} + \mybar{\eta}\ts{c} + \mybar{\Delta \Phi}\ts{Elec} + \mybar{\Delta \Phi}\ts{Solid}.
\end{equation}
When the OCV is sufficiently linear (see \cite{Marquis:2019} for more details), we have that 
\begin{equation}
    \mybar{U}\ts{eq}\left(c\ts{s,n},c\ts{s,p}\right) \approx U\ts{eq}\left(\mybar{c}\ts{s,n}, \mybar{c}\ts{s,p}\right)
\end{equation}
where the error introduced by making this approximation is of the same order of magnitude as the error introduced by the other approximations employed in the reduction of the DFN model. In contrast, if a particular point is chosen to evaluate the voltage then the error introduced is greater in magnitude. 

The remaining terms in (\ref{eqn:electrode-averaged-voltage}) are given by simple algebraic expressions obtained by employing the electrode-averaged exchange current in the current equations of the DFN model instead of the Butler--Volmer expression. The electrode-averaged reaction overpotentials are given by
\begin{equation}
    \mybar{\eta}\ts{r} = \mybar{\eta}\ts{r,p} - \mybar{\eta}\ts{r,n},
\end{equation} 
\begin{equation}
    \mybar{\eta}\ts{r,k} =-\frac{2RT}{F}\sinh^{-1}\left(\frac{I}{a\ts{k}\mybar{j}\ts{$0$,k} L\ts{k}}\right), \quad \text{k}=\{\text{p},\text{n}\},
\end{equation} 
with the electrode-averaged exchange-current densities given by
\begin{equation}
	 \mybar{j}\ts{$0$,k} = m\ts{k}  (c\ts{s,k})^{1/2}(c\ts{s,k,max}-c\ts{s,k})^{1/2} (\mybar{c}\ts{e,k})^{1/2}.
\end{equation}
The electrode-averaged electrolyte concentration overpotential is given by
\begin{equation}
	 \mybar{\eta}_c =  \frac{2RT}{F c\ts{e,typ}} (1-t^+)\left(\mybar{c}\ts{e,p} - \mybar{c}\ts{e,n}\right),
\end{equation}
and the electrode-averaged electrolyte Ohmic losses are given by
\begin{equation}
   \mybar{\Delta \Phi}_{\text{Elec}}= -\frac{I}{\kappa\ts{e,typ}}\left(\frac{L\ts{n}}{3\epsilon\ts{n}^b} + \frac{L\ts{s}}{\epsilon\ts{s}^b} + \frac{L\ts{p}}{3\epsilon\ts{p}^b} \right).
\end{equation}
Finally, the electrode-averaged solid-phase Ohmic losses are given by
\begin{align}
	 &\mybar{\Delta \Phi}_{\text{Solid}} =  -\frac{I}{3}\left(\frac{L\ts{p}}{\sigma\ts{p}} + \frac{L\ts{n}}{\sigma\ts{n}} \right).
     \end{align} 
\end{subequations}

\begin{table*}[]
 \captionsetup{width=\textwidth,justification=justified,singlelinecheck=false}
	\centering 
	\resizebox{\textwidth}{!}{%
	\begin{tabularx}{\textwidth}{c c l c c c} 
	\toprule
     Parameter & Units & Description & $\Omega\ts{n}$ & $\Omega\ts{s}$ & $\Omega\ts{p}$ \\ 
    \midrule 
    $\epsilon\ts{k}$ & - & Electrolyte volume fraction & $0.3$ & 1 & $0.3$  \\
    $c\ts{k,max}$ & $\SI{}{\mol / \metre^{3}}$ & Maximum lithium concentration &$2.4983\times 10^4$ & - & $5.1218\times 10^4$ \\
     $\sigma\ts{k}$ & $\SI{}{ \siemens / \metre}$ & Solid conductivity & 100 & - & 10 \\
     $D\ts{s,k}$ & $\SI{}{\metre^{2}/\second}$ & Electrode diffusivity &  $3.9\times10^{-14}$ & - &  $1\times10^{-13}$\\
     $R\ts{k}$ & $\SI{}{\micro\metre}$ & Particle radius & 10 & - & 10  \\
     $a\ts{k}$ & $\SI{}{\micro\metre^{-1}}$ & Electrode surface area density & 0.18 & - & 0.15 \\
     $m\ts{k}$ & $\SI{}{(\ampere / \metre^{2})(\metre^3/\mol)^{1.5}}$ & Reaction rate & $2\times10^{-5}$ & - & $6\times10^{-7}$ \\
     $L\ts{k}$ & $\SI{}{\micro\metre}$ & Thickness & 100 & 25 & 100 \\
     $U\ts{k,ref}$ & $\SI{}{\volt}$ & Reference OCP& 0.18 & - & 3.94 \\
      \midrule
    $c\ts{e, typ}$ & $\SI{}{\mol / \metre^{3}}$ & Typical lithium-ion concentration in electrolyte & & $1\times10^3$ & \\
    $D\ts{e,typ}$ & $\SI{}{\metre^{2}/\second}$ & Typical electrolyte diffusivity &  & $5.34\times10^{-10}$ &  \\
    $\kappa\ts{e,typ}$ & $\SI{}{\siemens/\metre}$ & Typical electrolyte conductivity &  & $1.1$ &  \\
     $F$ & $\SI{}{\coulomb / \mol}$ & Faraday's constant & & 96485 &  \\
     $R$ & $\SI{}{\joule / (\mol \, \kelvin)}$ & Universal gas constant && 8.314472 & \\
     $T$ & $\SI{}{\kelvin}$ & Temperature && 298.15 & \\
     $b$ & - & Bruggeman coefficient && 1.5 &  \\
     $t^+$ & - & Transference number && 0.4 & \\
     $I\ts{typ}$ & $\SI{}{\ampere / \metre^2}$ & Typical current density & & 24 ($\SI{1}{C}$)& \\
    \bottomrule
	\end{tabularx}}
	\vspace{0.25cm}
    \caption{Model parameters with values taken from \citep{SMouraGithub}.} \label{tab:parameters}
\end{table*} 

\section{Bayesian parameter estimation} 

\subsection{Comparison with frequentist approach}

The Bayesian approach to parameter estimation differs from the frequentist approach by allowing probability distributions over all possible parameter values to be calculated.  Statistics such as mean, variance, etc.\ are then computed directly from these distributions.

By contrast, in the frequentist school, parameters are assumed fixed (but unknown) and estimators are constructed to for them, such as MLE. The variance may be quantified by calculating the Cram\'er-Rao lower bound (CRLB) at the MLE point. This involves taking the inverse of the Fisher information matrix (FIM), defined as the variance of the score (gradient of the log-likelihood with respect to parameters) at the optimum point. This is also equal to the expectation of the negative Hessian of the likelihood function, so that
\begin{equation}
\label{eqn:FI}
    \bar{I}(\theta) = \mathbb{E}\left[\left(\frac{\partial L}{\partial\theta}\right)^2\right] = -\mathbb{E}\left[\frac{\partial^2 L}{\partial\theta^2}\right],
\end{equation}
where $L = \log p(y|\theta)$ ($\theta$ scalar). Alternatively, for a functional of the parameter $\theta \mapsto f(\theta)$ ($\theta$ vector), its information is given by
\begin{equation}
    \label{eqn:obs_FI}
    I(\theta) = \frac{\partial f}{\partial\theta}^T \bar{I}(f(\theta))\frac{\partial f}{\partial\theta},
\end{equation}
which provides a simple numerical evaluation method as $I(f(\theta))$ becomes the inverse of the variance for a Gaussian likelihood function \citep{Tangirala2015}.
The Cram\'er-Rao lower bound for variance is then retrieved by inverting the FIM calculated at the maximum likelihood point,
\begin{equation}
\label{eqn:crlb}
  \Sigma\ts{CRLB}(\hat{\theta}\ts{MLE}) = I(\hat{\theta}\ts{MLE})^{-1}.
\end{equation}

 Critically this method only applies at the MLE point and its infinitesimal neighbourhood. In the asymptotic case with enough observations, this method of estimating parameter uncertainty will converge with the Bayesian method, which performs the uncertainty estimation by calculating the variance of the posterior distribution. This is the result of the posterior parameter distribution tending to a Gaussian \citep{Ghosh2006}. However, for the non-asymptotic case, where the posterior may be more complex, this frequentist measure may not reflect the true covariance of the parameter vector.

\subsection{Bayesian approach}
We wish to estimate the posterior probability distribution of the parameter vector $\theta$ conditional on the observed data $y$. According to Bayes' rule, this is given by

\begin{equation}
\label{eqn:BayesRule}
    p(\theta|y) =\frac{p(y|\theta)p(\theta)}{p(y)} .
\end{equation}

When there is no closed form solution to the posterior distribution, it can be approximated numerically. Markov Chain Monte Carlo (MCMC) methods achieve this by constructing a Markov chain which has the posterior as its stationary distribution. Following that, if point estimates of the parameter are required, they may be retrieved by evaluating properties of the posterior distribution.
The type of point estimate required depends on the loss function chosen. To minimise the mean squared error over the posterior, the solution is the mean of the posterior distribution \citep{Sarkka2013}, namely
\begin{equation}
\label{eqn:expect_int}
    \hat{\theta}\ts{MMSE} = \int_\theta \theta p(\theta|y) d\theta.
\end{equation}
Given a Markov chain of the parameter vector, this integral may be approximated by simply taking the arithmetic mean of its stationary state \citep{Gilks1996}. The significant advantage of using MCMC to approximate the posterior distribution is that only the unnormalized posterior probability needs to be calculated during each iteration, meaning there is no need to estimate the denominator in (\ref{eqn:BayesRule}).

For the purposes of parameter estimation, we consider the robust adaptive Metropolis-Hastings (RAMH) MCMC algorithm \citep{Vihola2012}. It is an extension of the standard random walk Metropolis-Hastings recursion \citep{Metropolis1953,Hastings1970}, which constructs a Markov chain by recursively drawing candidate vectors $\theta_c$ from a proposal distribution conditional on the current vector $\theta_t$, $q(\theta_c|\theta_t)$, which is assumed symmetric, so that $q(\theta_c|\theta_t)=q(\theta_t|\theta_c)$. 
For a random walk in parameter space, we use the proposal distribution \citep{Sarkka2013}
\vspace{0.2cm}
\begin{equation}
\theta_c \sim N(\theta_t,\Sigma), 
\end{equation}
a normal distribution with mean $\theta_t$ and covariance $\Sigma$. The acceptance probability of the candidate vector $\theta_c$ is then  calculated by the ratio of posterior probabilities between $\theta_c$ and $\theta_t$. The adaptive aspect enables the RAMH to overcome the challenge in designing the proposal distribution (i.e., matrix $\Sigma$). For efficient sampling, the design should be such that the chain convergences to its stationary state at an acceptable rate while sampling efficiently allows exploring around the stationary distribution. It has been shown that a suitable acceptance rate for this purpose for the RWMH is $\bar{\alpha}=0.234$ \citep{Roberts2001}. The RAMH algorithm attempts to achieve the chosen target acceptance rate $\alpha^*$ by adjusting the proposal distribution covariance matrix based on previous observations as shown in Table \ref{tbl:RAMH}.

\begin{table}[]
\captionsetup{width=\columnwidth,justification=raggedright,singlelinecheck=false}
\centering
\normalsize
\renewcommand{\arraystretch}{1.5}
\begin{tabularx}{\columnwidth}{c l}
 \Xhline{0.8pt}
\multicolumn{2}{l}{Initialization:} \\
& $\qquad \theta\ts{t} = \theta\ts{0} ~,~ \ SS^T= \Sigma\ts{0}$ \\
\multicolumn{2}{l}{Sample $\theta\ts{c}$:} \\
& $\qquad \theta\ts{c} = \theta\ts{t} + Sw ~,~ w \sim N(0,1)$ \\
\multicolumn{2}{l}{Evaluate acceptance probability $\alpha$:} \\
& $\qquad  \alpha = \min \left\{1, \frac{p(\theta\ts{c}|y)}{p(\theta\ts{t}|y)} \right\}$ \\
\multicolumn{2}{l}{Accept candidate $\theta\ts{c}$ with probability $\alpha$:} \\
& $\qquad  \textrm{if} ~ \alpha > U(0,1) ~,~ \theta\ts{c} = \theta\ts{t}$ \\     
\multicolumn{2}{l}{Update proposal covariance matrix:} \\
& $\qquad  SS^T = S\left(I + n^{-\gamma}(\alpha - \alpha^*)\frac {ww^T}{||w||^2}\right)S^T$ \\
 \Xhline{0.8pt}
\end{tabularx}
\vspace{0.2cm}
\caption{Robust adaptive Metropolis-Hastings algorithm \citep{Vihola2012}, where $n$ is the iteration number and $\gamma$ a parameter determining the speed of adaptation of the proposal density covariance matrix. $S$ in the update step may be obtained by Cholesky decomposition.}
\label{tbl:RAMH}
\end{table}

\section{Numerical Results}

\subsection{SPMe Numerical Implementation} 
The SPMe is relatively computationally cheap to run, but efficient numerical methods are still required to facilitate the many thousands of model runs required by the Bayesian parameter estimation approach outlined in this paper. To this end, we employed spectral collocation to discretize the spatial dimensions in the model, using Chebyshev polynomials as the basis functions and Chebyshev nodes for the collocation points, in a similar approach to that pioneered for the DFN model by \citep{bizeraythesis, bizeraySPECTRAL}. This greatly reduced the number of states in the system compared to standard discretization approaches such as finite difference or finite volume. 

After spatial discretization, an ODE system is obtained. To further speed up its numerical solution, (\ref{eqn:SPMe-particles}) and (\ref{eqn:SPMe-electrolyte}) were converted to discrete time form using MATLAB's  \texttt{c2d} function. These enhancements enabled a single full constant current discharge to be performed in \SI{10}{ms} when using three basis functions in each particle, and eight in the electrolyte. The discrete time approach also allowed us to achieve comparable solution times for non-constant input currents.

\subsection{Implementation of Bayesian approach}

To explore the identifiability of the SPMe using the RAMH method, we focused on four parameters consisting of the diffusivities in both particles and the electrolyte, $D\ts{n}, D\ts{p}, D\ts{e}$, and the transference number $t^+$. In addition we  estimated the variance of the output noise, $\sigma^2$. All other parameters were considered known including the open circuit potential functions. We considered two different excitations. Firstly, we simulated the voltage response to a zero-bias multiharmonic sinusoidal excitation. The excitation had an amplitude of \SI{8}{mV} with 4 harmonics, once a decade between \SI{100}{mHz} and \SI{100}{Hz}. This excitation was applied at 11 different equispaced SoC levels and sampled at \SI{4000}{Hz}. Following that, we used a \SI{1}{mHz} C/24 single harmonic sinusoidal signal with 1C DC bias to achieve a broad excursion in SoC. In each case we added independently and identically distributed (i.i.d.) measurement noise to the voltage response with variance $\sigma^2$ corresponding to a 1\% two-sigma error in the amplitude of the voltage response. For the MCMC analysis,  the three diffusivities $D\ts{n},D\ts{p},D\ts{e}$ were first scaled by multiplying them by 10\textsuperscript{14}, 10\textsuperscript{13} and 10\textsuperscript{10} respectively, to bring them in the range [0,10].  The logarithm of the noise variance was taken in order to bring it within the same order of magnitude. The priors were chosen by
\begin{equation*}
    p(\theta) = 
    \begin{cases}
        \Gamma(k,s),  & \theta \in \{ D\ts{n}, D\ts{p}, D\ts{e}\}, \\
        \text{Beta}(\alpha,\beta),  & \theta \in \{t^+ \}. \\
    \end{cases}
\end{equation*}
The shape and scale parameters $k$ and $s$ for the diffusivity priors were chosen so that the mode of the distribution was at the real value and that the 99\% cumulative probability lay at a value of 100 on the new scale. For the transference number, $t^+$, we chose $\alpha=4$ and $\beta=5.5$ to obtain a mode of 0.4 and 80\% of probability mass between 0.2 and 0.6. For the system measurement noise, a uniform distribution on an infinite interval was used.

The initial $\theta\ts{c}$ for the Markov Chain was drawn at random so that $p(\theta_c)>0$. The initial proposal distribution covariance matrix was set as $\Sigma_0 = 0.001I$, where $I$ is the identity. The likelihood function was chosen to be Gaussian,
\begin{equation*}
    p(y|\theta) \sim N(f(\theta),\sigma^2),
\end{equation*}
where $f(\theta)$ is given by the SPMe and $y$ is the voltage output, giving the logarithm of the unnormalized posterior probability for $\theta$ by
\begin{equation}
\label{eqn:posterior_prob}
    \log p(\theta|y) = \log p(\theta) -\frac{n}{2} \log (2\pi\sigma^2) - \sum_{i=1}^n\frac{(y-f(\theta))^2}{2\sigma^2},
\end{equation}

where $n$ is the sample size. The RAMH method evaluates the exponential of (\ref{eqn:posterior_prob}) in the recursion (Table \ref{tbl:RAMH}) to achieve the MCMC approximation of the posterior distribution. The total number of MCMC iterations in each case was 100,000, with a 10,000 iteration burn-in period. Following that, we calculated the MMSE estimator for the posterior distribution in each case. In addition to MCMC, in each case we performed MLE and calculated the observed Fisher information to obtain the CRLB estimates for comparison. The initial guesses for MLE were random but constrained to within 10\% of the true values so that the CRLBs calculated were at a local optima near the true values.

\subsection{Results}

\begin{figure*}[thpb]
  \centering
 \begin{subfigure}[b]{0.49\textwidth}
 \caption{}
  \hskip-0.6cm
  \vskip-0.6cm
 \includegraphics[width=9.43cm,height=7.5cm]{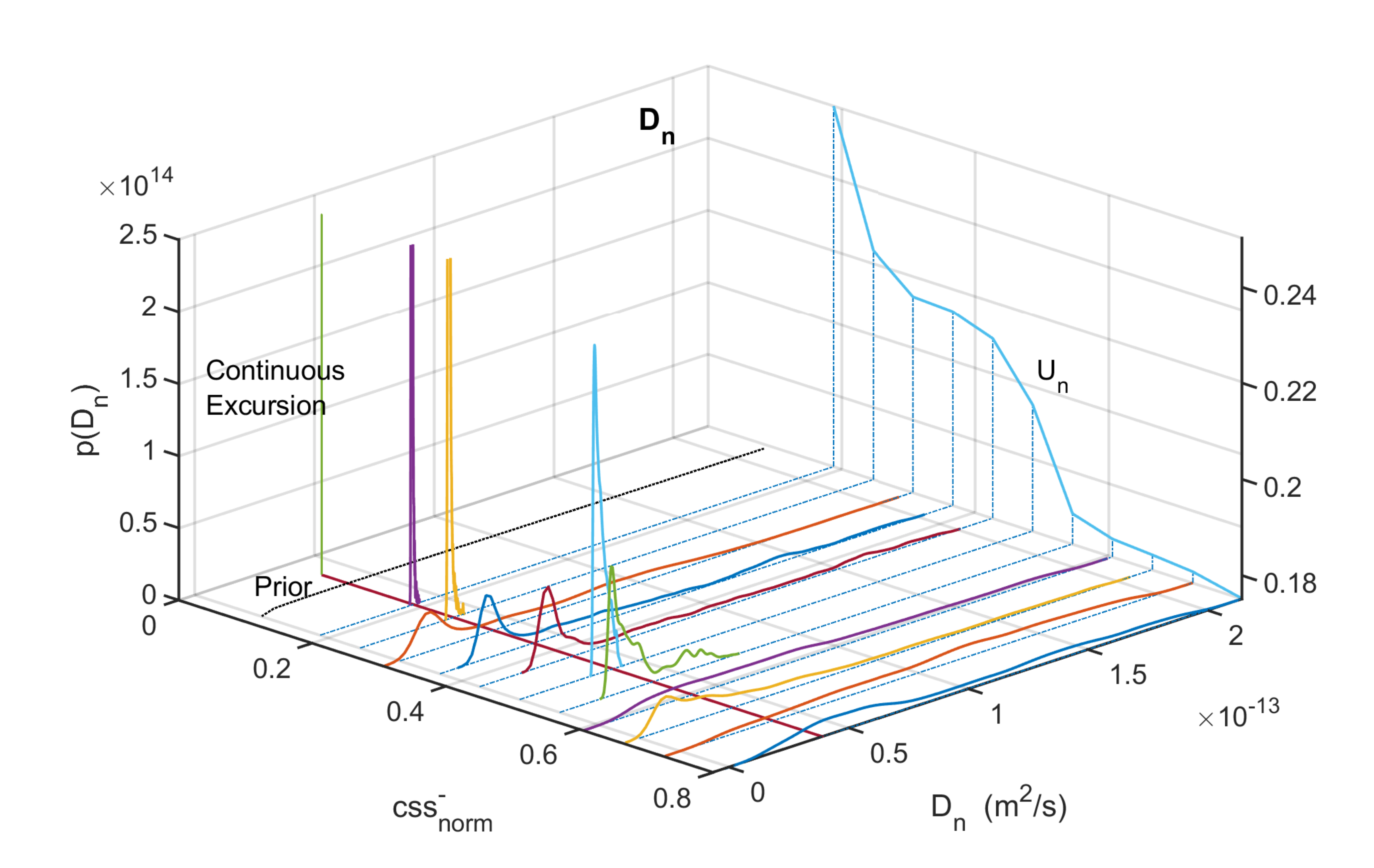}
 \label{fig:Dn}
 \vspace{-0.1cm}
 \end{subfigure}
 \begin{subfigure}[b]{0.49\textwidth}
 \hskip-0.6cm
 \vskip-0.6cm
 \includegraphics[width=9.43cm,height=7.5cm]{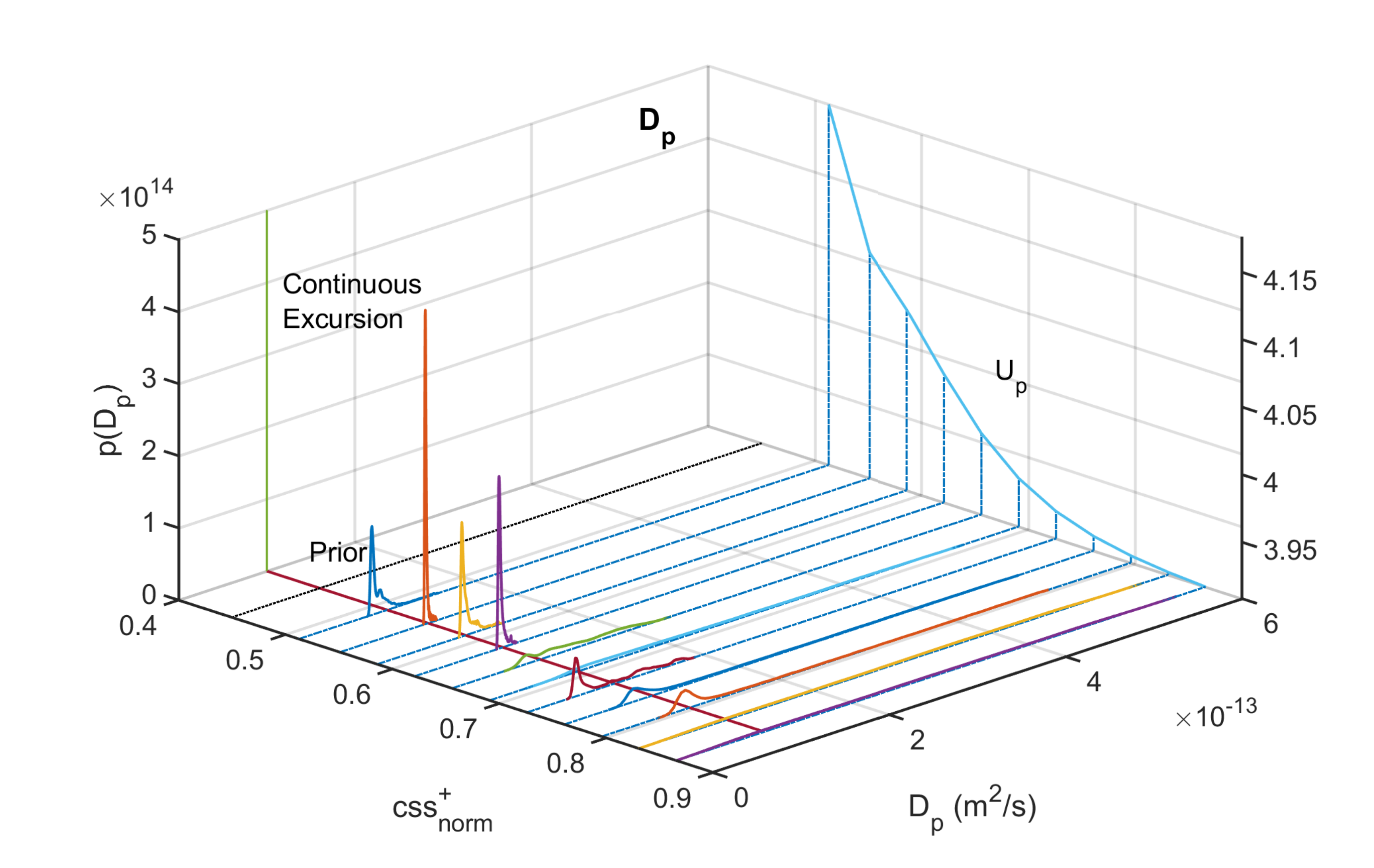}
 \label{fig:Dp}
 \vspace{-0.1cm}
\end{subfigure}
 \centering
 \begin{subfigure}[b]{0.49\textwidth}
 \hskip-0.6cm{\tiny }
  \vskip-0.6cm
 \includegraphics[width=9.43cm,height=7.5cm]{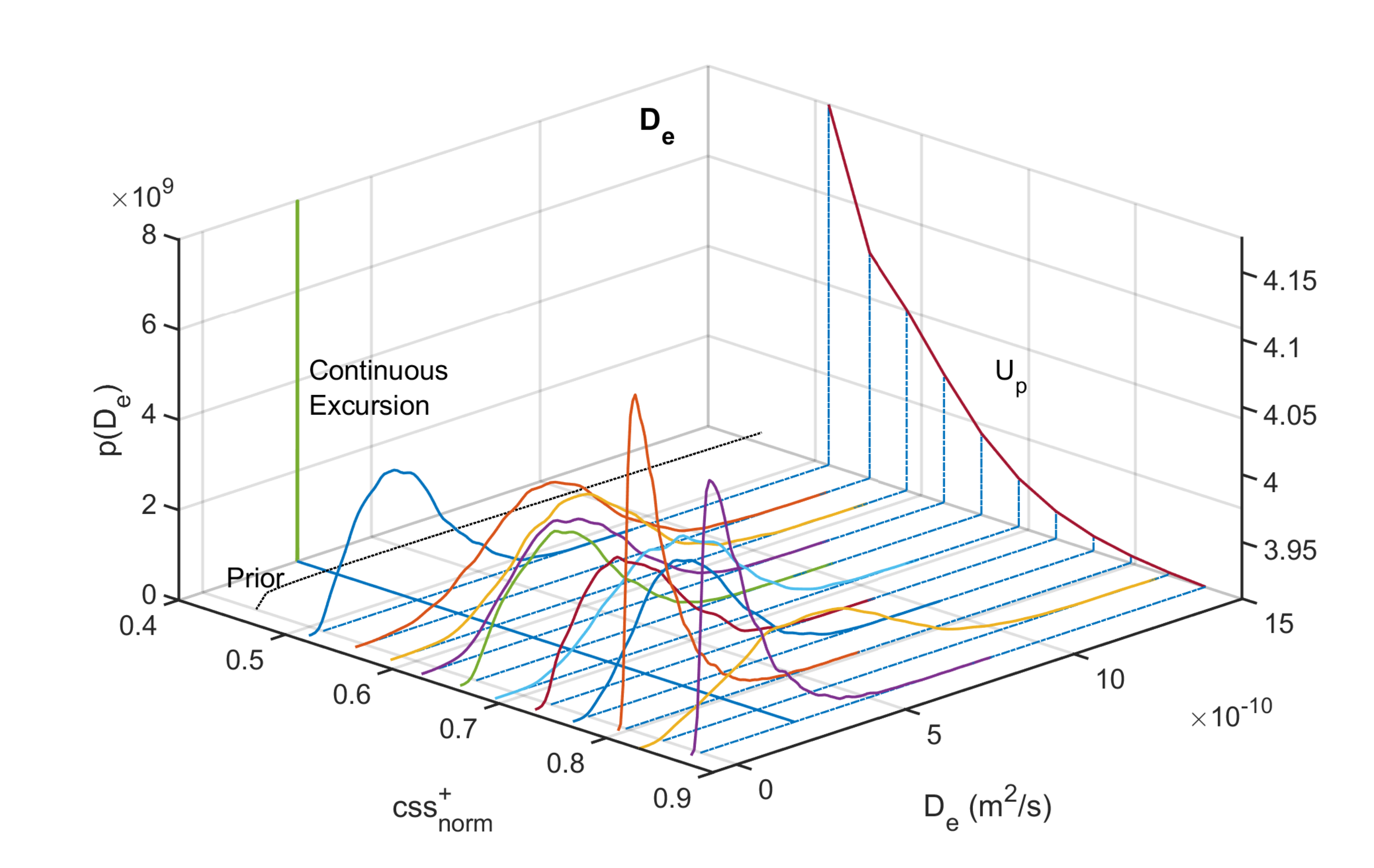}
 \label{fig:De}
  \vspace{-0.1cm}
 \end{subfigure}
 \begin{subfigure}[b]{0.49\textwidth}
 \hskip-0.6cm
  \vskip-0.6cm
 \includegraphics[width=9.43cm,height=7.5cm]{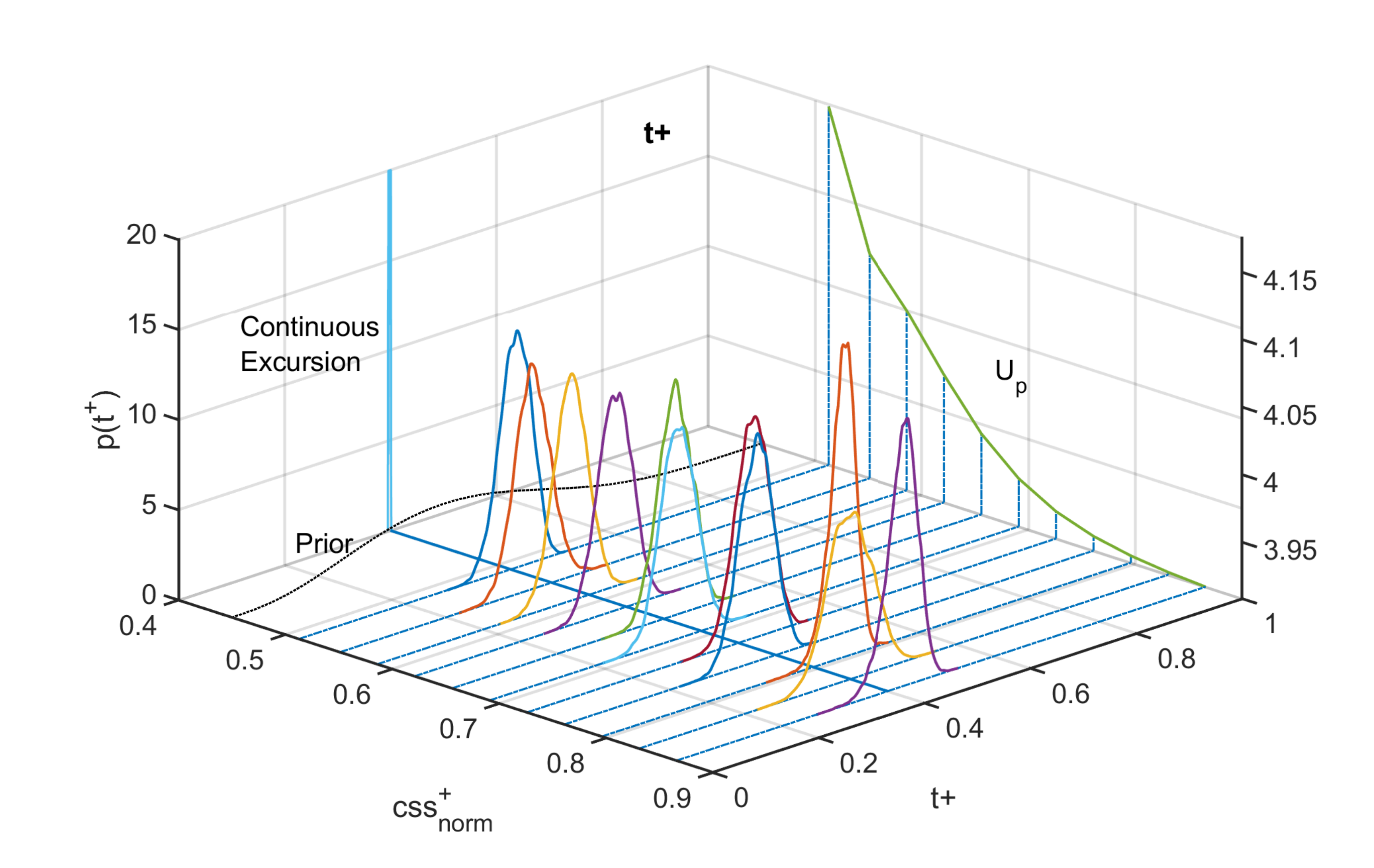}
 \label{fig:tplus}
  \vspace{-0.1cm}
  \hspace{-0.6cm}
 \end{subfigure}
 \vspace{-0.4cm}
 \caption{Marginal posterior densities for the four parameters: $D\ts{n}$, $D\ts{p}$, $D\ts{e}$, $t^{+}$ for different SoCs. The x-axes show the normalised negative and positive electrode Li-ion surface concentrations at each SoC. The equivalent point on the half cell OCP of the positive/negative electrode is highlighted by the dashed lines. The horizontal line in the x-y plane perpendicular to these is the true value for the parameter. The posteriors for the wide excursion are shown on the left edge, accompanied by the priors. The full range of the wide excursions posteriors is not shown so that the local cases remain clearly visible.}
\label{fig:case_1}
\end{figure*}

Our simulations illustrate the contrast in identifiability, in the chosen parameter set, using zero bias-local excitation signals with respect to an excitation over a wide SoC excursion. Fig. \ref{fig:case_1} depicts the marginal posterior distributions for the four model parameters. It is clear that in the case of the solid state diffusivities $D\ts{n}$ and $D\ts{p}$, the identifiability is dependent on the gradient of the respective half cell OCP at the point of excitation. This behaviour is expected \citep{Bizeray2018}, as the two parameters affect the transient behaviour of surface Li-ion concentration on each electrode. As a result of the small gradient of the negative electrode at most SoC values, $D\ts{n}$ has poor identifiability across most of the SoC range considered. By contrast, the identifiability of the electrolyte diffusivity $D\ts{e}$ and transference number $t^+$ are less sensitive to the SoC at point of excitation. 

\begin{figure*}[h]
\centering
\begin{subfigure}[b]{0.49\textwidth}
\caption{}
\vspace{-0.25cm}
\hspace{-1.5cm}
\includegraphics[width=11cm,height=9.25cm]{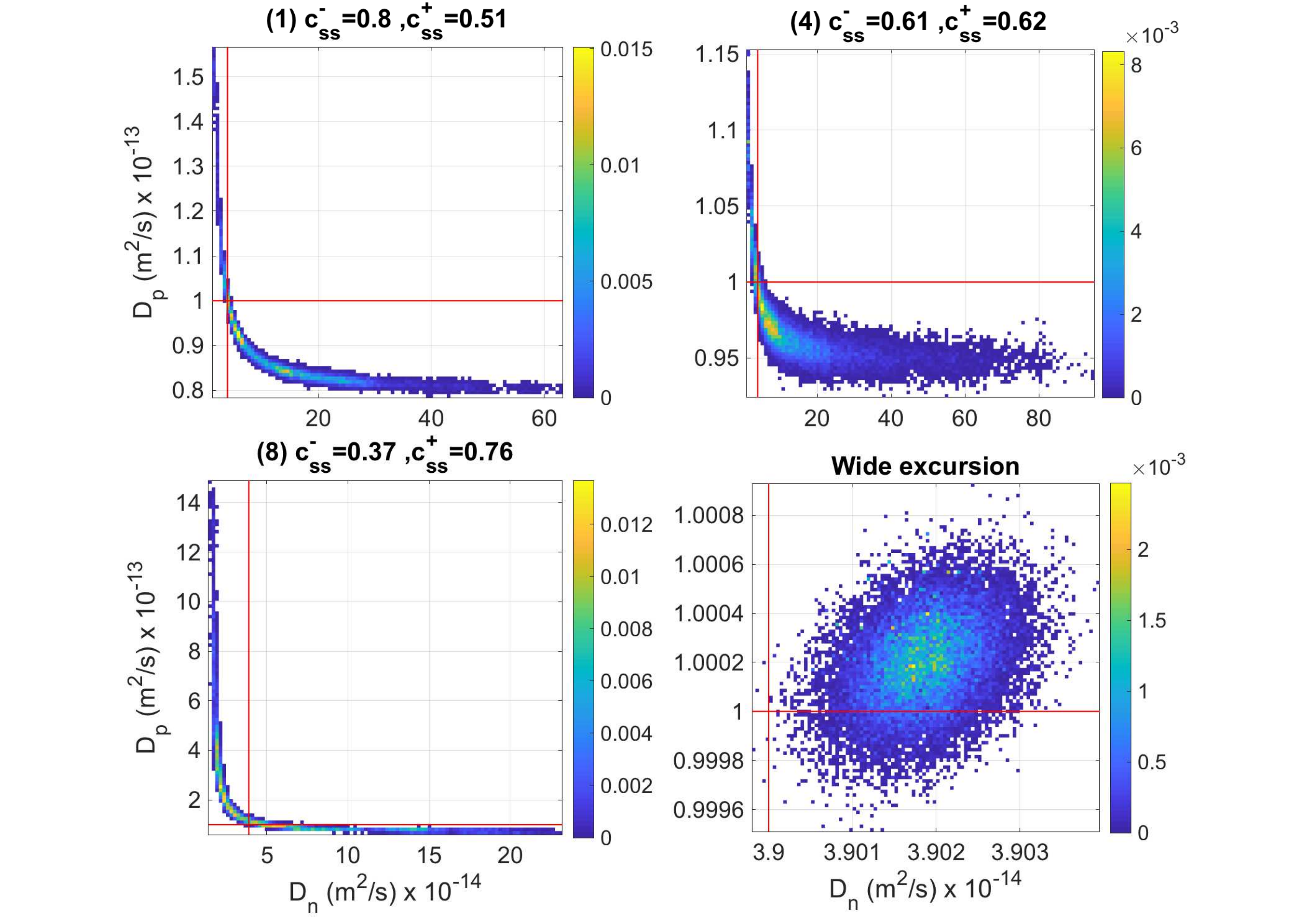}
\label{fig:DnDp}
\vspace{-0.5cm}
\end{subfigure}
\begin{subfigure}[b]{0.49\textwidth}
\caption{}
\vspace{-0.25cm}
\hspace{-0.6cm}
\includegraphics[width=10.5cm,height=9.2cm]{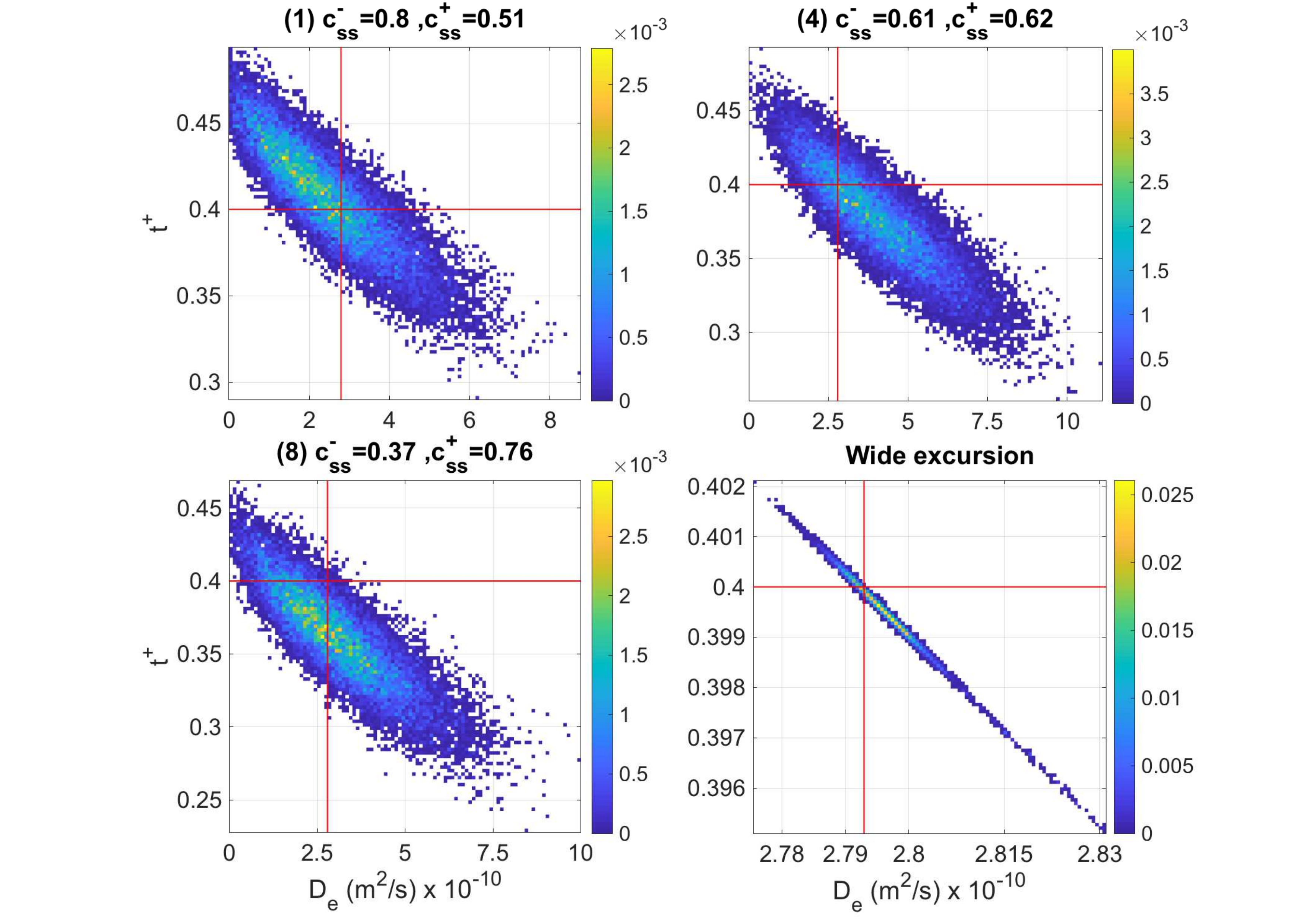}
\label{fig:Detplus}
\vspace{-0.5cm}
\end{subfigure}
\caption{MCMC joint posterior probability distributions (a) $p(D\ts{n}, D\ts{p})$ and (b) $p(D\ts{e}, t^+)$ , for a subset of local excitation points and the wide excursion. The red crosses highlight the true parameter values.}
\label{fig:3d_histos}
\end{figure*}

It can be seen in Fig. \ref{fig:case_1} and \ref{fig:3d_histos} that the marginal posterior distribution from the excitation with wide SoC excursion is much tighter for all parameters. Moreover, it is clear that the posterior distributions approach the asymptotic case with the wide SoC excursion and the variances calculated by the Bayesian and frequentist methods converge. In contrast, for local excitations, the posterior distributions of all the diffusivities are asymmetric with a long tail. This is particularly pronounced for $D\ts{n}$. This is due to the loss of sensitivity of the likelihood function at high diffusivity values, as the transient response becomes fast enough so as not to impact the voltage response. Also, the prior used was such that the 99\textsuperscript{th} percentile lies at 100 on the scaled parameter value, meaning that the relatively large variance of the prior weakly constrains the posterior probability distribution. For the transference number $t^+$, the posteriors are symmetric and have low variance, making the MLE and MMSE estimates and uncertainties similar across all cases, with low variances. The estimation of the system measurement noise is very similar across local and global cases regardless of method used. 

For the diffusivities, in 29 out of 33 local cases, the Cram\'er-Rao bound is higher than the posterior variance from MCMC. This is most obvious in the poorly identifiable cases for $D\ts{n}$, using SoC points 3,4 and 9. There are two factors contributing to the differences in the parameter variance estimates between the frequentist and Bayesian methods. Firstly, the local curvature of likelihood does not reflect the global variance, which is clear in the case of diffusivities which have heavily asymmetric posteriors. Secondly, prior information, which is not considered by the frequentist method, serves to reduce the variance in poorly identifiable cases.

A key advantage of obtaining an approximation of the posterior distributions comes from analysing the \textit{joint} posteriors as well as the marginals. In figure \ref{fig:3d_histos}, joint posterior distributions $p(D\ts{n},D\ts{p})$ and $p(D\ts{e},t^+)$ for a subset of excitation points are illustrated. For the wide excursion case, where the marginal posteriors have variances similar to the CRLB, the joint posteriors are clearly unimodal. However, the local excitation cases, for $D\ts{n},D\ts{p}$, it is clear that the posterior distribution implies the existence of multiple solutions, as the distribution has a poorly defined mode along the two dimensions. This effect is not captured by MLE and CRLB, which assume a unimodal distribution as a consequence of being a local measure.

\begin{table*}[t]
\captionsetup{width=\textwidth,justification=raggedright,singlelinecheck=false}
\centering
\begin{tabular}{|c|c|c|c|c|c|c|c|c|c|c|c|c|c|}
\hline
Parameter &  &\multicolumn{11}{c}{Excitation point} & \\
\hline
& & 1 &2 &3 &4 &5 &6 &7 &8 &9 &10 & 11 & Wide Excursion\\
\hline
& $c\ts{ss}^-$ & 0.80 & 0.73 & 0.67 & 0.61 & 0.55 &
0.49 & 0.43 & 0.37 & 0.31 & 0.25 & 0.19 & \\
& $c\ts{ss}^+$ & 0.51 & 0.55 & 0.59 & 0.62 & 0.66 &
0.69 & 0.73 & 0.76 & 0.80 & 0.83 & 0.87 & \\
\hline
$D\ts{n}$ & $\theta\ts{MMSE}$ & 13.98& 19.47& 13.24& 14.81& 3.27& 3.39& 12.74& 5.92& 5.59& 3.73& 3.83& 3.90 \\ 
& $\sigma\ts{MCMC}$ & 10.55& 14.64& 16.48& 14.41& 1.27& 0.27& 11.64& 5.01& 4.88& 0.08& 0.03& 5.05e-04 \\ 
& $\theta\ts{MLE}$ & 4.04& 4.23& 4.00& 4.08& 4.05& 4.09& 4.22& 4.04& 4.17& 3.90& 3.89& 3.90 \\ 
& $\sigma\ts{CRLB}$ & 15.29& 83.73& 91.69& 527 & 15.66& 4.82& 19.67& 22.66& 295 & 0.63& 0.16& 5.34e-04 \\ 
\hline
$D\ts{p}$ & $\theta\ts{MMSE}$ & 0.90& 0.98& 1.01& 0.98& 1.36& 2.72& 0.91& 2.28& 2.17& 22.41& 14.77& 1.00 \\ 
& $\sigma\ts{MCMC}$ & 0.12& 0.02& 0.08& 0.03& 0.42& 1.58& 0.40& 2.24& 2.09& 25.09& 14.28& 1.70e-04 \\ 
& $\theta\ts{MLE}$ & 1.00& 1.00& 0.99& 0.99& 0.96& 0.86& 0.89& 1.11& 1.01& 0.98& 1.01& 1.00 \\ 
& $\sigma\ts{CRLB}$ & 0.95& 0.91& 0.95& 6.56& 2.12& 2.63& 2.99& 4.07& 17.06& 2.65& 2.26& 1.71e-04 \\ 
\hline
$D\ts{e}$ & $\theta\ts{MMSE}$ & 2.43& 5.51& 5.34& 4.04& 2.73& 4.76& 2.49& 3.07& 0.98& 4.69& 1.19& 2.80 \\ 
& $\sigma\ts{MCMC}$ & 1.29& 1.79& 1.74& 1.59& 1.43& 1.68& 1.41& 1.40& 0.85& 2.20& 1.01& 6.61e-03 \\ 
& $\theta\ts{MLE}$ & 3.04& 2.86& 2.95& 3.00& 3.07& 2.88& 2.86& 2.82& 2.93& 3.01& 2.84& 2.80 \\ 
& $\sigma\ts{CRLB}$ & 1.63& 1.71& 1.70& 2.92& 1.89& 1.77& 1.64& 1.62& 3.09& 2.30& 2.29& 4.52e-03 \\ 
\hline
$t^+$ & $\theta\ts{MMSE}$ & 0.41& 0.36& 0.37& 0.38& 0.42& 0.36& 0.43& 0.36& 0.45& 0.40& 0.43& 0.40 \\ 
& $\sigma\ts{MCMC}$ & 0.03& 0.03& 0.03& 0.03& 0.03& 0.03& 0.03& 0.03& 0.02& 0.04& 0.03& 8.32e-04 \\ 
& $\theta\ts{MLE}$ & 0.40& 0.40& 0.41& 0.40& 0.41& 0.39& 0.38& 0.37& 0.42& 0.43& 0.40& 0.40 \\ 
& $\sigma\ts{CRLB}$ & 0.03& 0.03& 0.04& 0.03& 0.04& 0.04& 0.04& 0.04& 0.04& 0.05& 0.05& 5.72e-04 \\ 
\hline
$\sigma^2 \dagger$  & $\theta\ts{MMSE}$ & 1.60& 1.59& 1.60& 1.60& 1.61& 1.59& 1.61& 1.61& 1.58& 1.59& 1.61& 1.39 $\dagger\dagger$\\ 
& $\sigma\ts{MCMC}$& 0.11& 0.11& 0.11& 0.11& 0.11& 0.11& 0.11& 0.11& 0.11& 0.11& 0.11& 3.45 $\dagger\dagger$ \\ 
& $\theta\ts{MLE}$ & 1.60& 1.59& 1.60& 1.60& 1.61& 1.59& 1.62& 1.61& 1.58& 1.59& 1.61& 1.60 \\ 
& $\sigma\ts{CRLB}$ & 0.11& 0.11& 0.11& 0.11& 0.11& 0.11& 0.11& 0.11& 0.11& 0.11& 0.11& 0.11 \\ 
\hline
\end{tabular}
\vspace{0.25cm}
 \caption{Parameter and uncertainty estimates of the scaled variables using Bayesian and frequentist methods. $\theta\ts{MMSE}$ indicates the posterior mean and $\sigma\ts{MCMC}$ the posterior standard deviation. For MLE, $\theta\ts{MLE}$ is the point estimate and $\sigma\ts{CRLB}$ the standard deviation calculated from the Cram\'er-Rao lower bound. $\dagger$ scaling for $\sigma^2$ is 10\textsuperscript{-9} for mean and $\sigma$ estimates. $\dagger\dagger$ The MCMC for $\sigma^2$ converged slowly. Values consistent with the local cases were reached after 40'000 iterations.} 
 
\label{tbl:case_1}

\end{table*}

\section{Conclusion}

We have used a Bayesian framework with MCMC methods to analyse the identifiability of a subset of parameters of the SPMe model. The two solid state diffusivities, the electrolyte diffusivity and transference number show different characteristics with respect to identifiability when zero bias multi-harmonic sinusoidal excitation at a single SoC point is applied. We showed that the local identifiabilities of the solid state diffusivities, determined by the variances of the marginal posterior distributions, depend on the respective half cell OCP gradients at the point of excitation. The electrolyte diffusivity and transference number do not display this local sensitivity. Furthermore, we demonstrated that the common frequentist approach to estimating parameter uncertainty is unreliable for local excitations around a single SoC, because the local covariance is not a good approximation of the global behaviour. However, this problem is negligible in the case where we applied a bias to a sinusoidal pulse, giving a wide excursion over a range of SoCs. In this case, the parameter and variance estimates for the Bayesian and frequentist methods converge. This has significant implications for the case where parameters are assumed to be functions of SoC. If local estimates for parameters are retrieved experimentally at many SoC points and interpolated to approximate the parameter variation as a function of SoC, there is a significant risk of mis-identifying the function, because local identifiability may be poor.

\bibliography{IFAC_ref}           

\end{document}